\documentclass[runningheads]{llncs}
\usepackage[T1]{fontenc}
\usepackage{booktabs}
\usepackage{multirow}
\usepackage{adjustbox}
\usepackage{caption}
\usepackage{subcaption}
\usepackage{todonotes}

\usepackage{amsmath}

\DeclareMathAlphabet{\mathmybb}{U}{bbold}{m}{n}

\usepackage[pagebackref=true,breaklinks=true,colorlinks,bookmarks=false]{hyperref}

\usepackage{graphicx}
\usepackage[misc]{ifsym} 
\usepackage{color}

\begin{document}
\title{Prostate Lesion Estimation using Prostate Masks from Biparametric MRI}

%
\titlerunning{Prostate Lesion Estimation using Prostate Masks from Biparametric MRI}


\author{Ahmet Karagöz\inst{1}$^*$ , Mustafa Ege Seker\inst{2}$^*$,  Mert Yergin\inst{3}$^*$, Tarkan Atak Kan\inst{1}$^*$, Mustafa Said Kartal\inst{4}$^*$, Ercan Karaarslan\inst{2}$^*$, Deniz Alis\inst{2}$^*$ \and
Ilkay Oksuz\inst{1}$^*$}

\authorrunning{Karagöz A., Şeker ME., Yergin M. et al.}


\institute{Computer Engineering, Istanbul Technical University, Istanbul, Turkey \and Acibadem Mehmet Ali Aydinlar University School of Medicine, Istanbul, Turkey \and Department of Software Engineering and Applied Sciences, Bahcesehir University, Istanbul, Turkey \and Faculty of Medicine, Sivas Cumhuriyet University, Sivas, Turkey
}


%
%

\maketitle              
\begin{abstract}
Biparametric MRI has emerged as an alternative to multi-parametric prostate MRI, which eliminates the need for the potential harms to the patient due to the contrast medium. One major issue with biparametric MRI is difficulty to detect clinically significant prostate cancer (csPCA). Deep learning algorithms have emerged as an alternative solution to detect csPCA in cohort studies.  We present a workflow which predicts csPCA on biparametric prostate MRI PI-CAI 2022 Challenge with over 10,000 carefully-curated prostate MRI exams. We propose to to segment the prostate gland first to the central gland (transition + central zone) and the peripheral gland. Then we utilize these predcitions in combination with T2, ADC and DWI images to train an ensemble nnU-Net model. Finally, we utilize clinical indices PSA and ADC intensity distributions of lesion regions to reduce the false positives. Our method achieves top results on open-validation stage with a AUROC of 0.888 and AP of 0.732.

\end{abstract}
\section{Introduction}
Magnetic resonance imaging (MRI) plays an imperative role in prostate cancer (PCa) diagnostics, and the number of prostate MRI scans is expected to significantly increase as the recent evidence suggests that performing pre-biopsy prostate MRI in men with suspicion of PCa \cite{ahmed2017diagnostic}. The main objective of prostate MRI is to identify clinically significant PCa (csPCa) (i.e., $Gleason Score \ge \ $3 + 4) while sparring men with benign lesions or indolent prostate from unnecessary interventions or treatment. The prostate imaging-reporting and data system (PI-RADS) was introduced in 2012 and updated twice to standardize prostate MRI acquisition and interpretation \cite{turkbey2019prostate}. The up-to-date multi-parametric prostate MRI (mpMRI) protocol includes axial T2-weighted images, diffusion-weighted images with a high b-value, apparent diffusion coefficient (ADC) maps calculated from DWI and dynamic contrast-enhanced images. The alternative to prostate mpMRI is bi-parametric prostate, which omits the contrast-enhanced sequence, thereby eliminating the potential hazards from the contrast medium. Though the benefits of the PI-RADS have been well recognized over the years, prostate MRI still suffers from intra-reader and inter-reader differences and non-negligible amounts of false-positive and false-negative results \cite{sonn2019prostate}.

Deep learning (DL) has shown remarkable performance on a broad spectrum of medical imaging tasks in recent years, with prostate cancer diagnostics no exception. However, despite the promises of DL technology in PCa, most of the proposed DL models have been trained on small single-center proprietary data and were not publicly shared, hindering the reliability and wide-spread adaptation \cite{van2021artificial}. ProstateX challenge partially addressed this problem, yet it did not have the adequate data size to train and test DL models for prostate MRI effectively \cite{westphalen2020variability}. 

PI-CAI (Prostate Imaging: Cancer AI) is a new grand challenge encompassing over 10.000 carefully annotated prostate MRI scans \cite{mckinney2020international}. The main goal of the challenge is to allow researchers to design, train and test publicly available DL models in large-scale for identifying csPCa on bi-parametric prostate MRI. In this challenge, we used the state-of-the-art medical image segmentation model, nnU-Net, as the base model and made several contributions to improve its performance. 





\section{Method}
Studies have shown a relationship between prostate volume, benign and malignant pathologies \cite{heidler2018correlation}. In particular, the combination of prostate volume with prostate-specific antigen (PSA) can be used as a biomarker in csPCA detections \cite{nordstrom2018prostate}. PSA density (PSAd) which is the ratio between PSA and prostate volume, is an important biomarker in today's prostate cancer guidelines such as PI-RADS. Interestingly, it was shown that Transition zone PSA density (TZPSAd) is a more accurate indicator than PSAd in csPCA detection \cite{schneider2019comparison}. This shows that prostate cancers, which are located within the prostate areas that host histologically different structures, differ in terms of cancer incidence, prognosis, and outcomes. For these reasons, before creating the prostate lesion segmentation / detection models, the prostate gland segmentation model was created. The central gland (transition + central zone) and the peripheral gland were segmented after. 

\subsection{Prostate Gland Segmentation}
\begin{figure}[htp]
    \centering
    \includegraphics[width=11.5 cm]{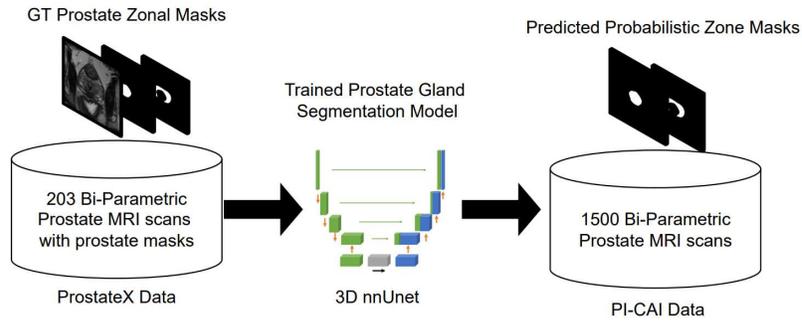}
    \caption{Prostate gland segmentation model}
    \label{fig:fig1}
\end{figure}

The prostate gland segmentation model was trained on 203 patients of the ProstateX \cite{cuocolo2021quality} dataset. Only the T2 modality was used in this model, because it was the modality in which anatomical structures can be best distinguished. nnU-Net is a self-configuring framework that selects the best architecture depending on the input dataset by adhering to a set of principles \cite{isensee2021nnu} (as detailed in Figure \ref{fig:fig1}). The sum of crossentropy and soft Dice loss is commonly applied by the nnU-Net framework, and the loss is applied at different resolutions (deep supervision). A standardized data preparation and augmentation pipeline is included with nnU-Net. In addition, extreme data augmentation techniques (such as Gaussian noise, Gaussian blur, brightness, contrast, simulation of low resolution and gamma augmentation, elastic deformation, scale, flip, mirror, and rotate) were used to maintain model performance at its highest level despite variations in acquisition parameters, resolution, image size, and quality. Five models, each a 3D nnU-Net model, were trained over the course of a thousand epochs using the training data that was split into five folds. On the PICAI dataset with 1500 Patients, it was found that there were no statistically significant differences between the performances of the single model and the ensemble of 5 models. In order to lessen the overall burden and shorten the prediction time, it was decided to employ a single model. 

\subsection{Data Preparation}
All images were converted from MHA (.mha) to nifti format (.nii.gz) as a pre-processing step for the lesion segmentation / detection models because nnU-Net requires nifti images within its own structure. To create images of the same size and resolution, each patient's ADC and DWI images were resampled, cropped, or padded. T2 was used as the primary modality within each patient, and the DWI and ADC images were adjusted to match T2 in terms of size and resolution. ADC images were normalized with complete z-score normalization with respect to the entire dataset while T2 and DWI images were normalized with instance-wise z-score normalization since ADC images are more robust than T2 and DWI images. Lesion segmentation was carried out using the nnU-Net framework, but instead of the default combination of dice and cross entropy loss, Focal Loss and cross entropy loss were employed, as in \cite{bosma2021report}. 

\subsection{Lesion Detection}
To begin with (as a first experiment), a nnU-Net model was trained using T2, ADC, and DWI images in order to establish a baseline model and track the effects of upcoming techniques on model performance. Throughout each experiment, the train and test splits were kept in order to completely remove the impact of changing data distribution on performance. Despite the dataset being divided into five folds, the initial experiments were only employed on one-fold. The main goal is to use the limited computational resources as effectively as possible while also reducing the workload required to train five models for each new method. Additionally, test-time augmentation was used to improve the model's performance by making predictions on many augmented images. To create a detection map that is required by the PICAI evaluation step, we acquired unique lesion candidates, as done in \cite{bosma2021report}, using the voxel-level confidence maps that were produced. By starting at the voxel with the highest degree of confidence and encompassing all related voxels (in 3D) with at least 40\% of the peak's degree of confidence, we specifically produced a lesion candidate. The candidate lesion is then eliminated from the softmax (confidence map), and the procedure is repeated up to the extraction of 5 lesions, if there are still remaining candidates. Small candidates with 10 voxels or fewer (0.009 cm3) are eliminated. 

\begin{figure}[htp]
    \centering
    \includegraphics[width=11.5 cm]{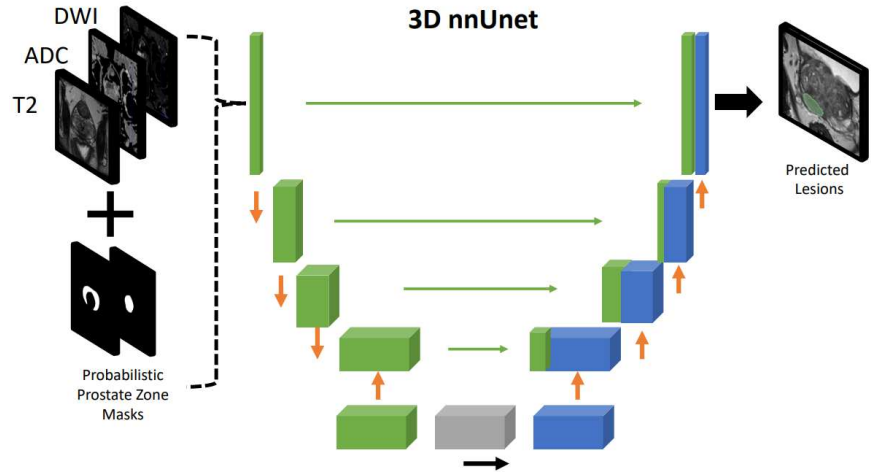}
    \caption{Prostate lesion segmentation model with prostate zone masks}
    \label{fig:fig2}
\end{figure}

\begin{figure}[htp]
    \centering
    \includegraphics[width=11.5 cm]{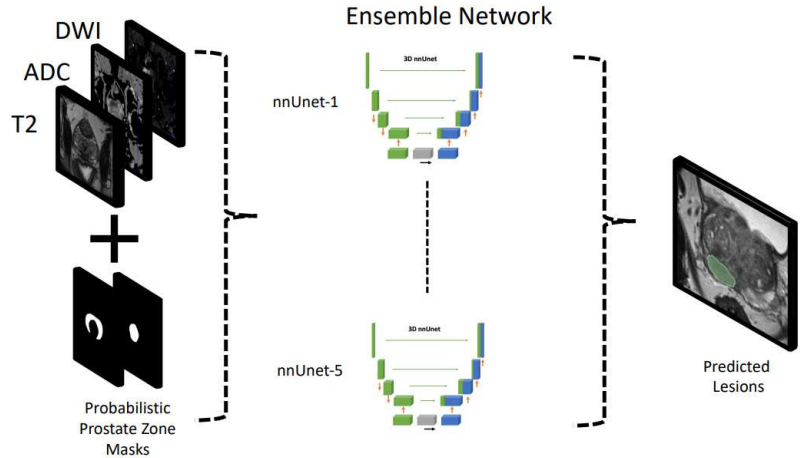}
    \caption{Ensemble of prostate lesion segmentation models}
    \label{fig:fig3}
\end{figure}

In the second experiment, the prostate gland segmentation model's output, Peripheral and Central gland predictions (probabilistic), were given to the model as separate modalities in addition to T2, ADC, and DWI (as detailed in Figure \ref{fig:fig2}). By doing this, the effects of prostate glands on lesion segmentation performance were examined. The third experiment did not involve any model training. Prediction performance of the model that had been trained in the second experiment were evaluated on cropped images. In this experiment we center cropped overly large pictures to remove the tissues that did not contain the prostate area in order to evaluate the effects of these deleted tissues on model performance. The largest and most dislocated prostatic regions are contained inside the cropped images of 81 mm x 192 mm x 192 mm size. In the fourth experiment, models were trained for five folds using T2, ADC, and DWI images, just as in experiment 1.  Model performances were evaluated using fivefold cross validation, and an ensemble of five models was tested on 100 patients  (PICAI open validation dataset) as detailed in Figure \ref{fig:fig3}. To assess how the models performed on cross validation and as an ensemble of five models on the test set, experiment 5 employed the same setup as experiment 2 for 5 folds. The sixth experiment did not involve any model training, but center-cropped images were used to evaluate the prediction capacities of models that were trained in experiment 5 (similar to experiment 3). 

\subsection{False positive elimination with clinical markers}
To determine if the provided clinical information is useful for csPCA detection or not, PSAd thresholding was performed in addition to the analysis carried out in the first six experiments using only images. PSAd was derived by dividing the PSA value from the clinical report by the prostate volume, which was determined using the segmentation model's predictions for the prostate gland. Although the probability of having csPCA is low in those with PSAd below the threshold value of 0.1 in a study \cite{girometti2022comparison}, the threshold value was determined as 0.04 in this experiment, thus reducing the possibility of false negative predictions. In other words, in Experiment 7, in addition to the setup in Experiment 6, PSAd thresholding was performed and those with PSAd value below 0.04 were directly counted as non-csPCA. Another point to emphasize is the intensity distribution of the lesions in the ADC images. As previous studies have shown \cite{saha2021end}, the ADC intensity distributions of lesions differ between benign and malignant cancers, so this difference can be used as a significant variable for lesion differentiation. In the final experiment (experiment 8), we investigated whether performance could be improved by thresholding the ADC intensity distributions of lesions. First of all, the ADC intensity distributions of the lesions of csPCA patients in the train dataset were examined to find the ideal threshold value. While deciding the threshold value as a result of histogram analysis, it is aimed to prevent possible false negative errors by adding a certain margin value and as a result, the value of 1200 was determined. Since it was not reasonable to use the lowest ADC intensity value in the predicted lesions as the threshold value, the threshold was made by calculating the average of the 10\% percentile. According to this threshold value, those with a value above 1200 have been determined as non-csPCA for FP elimination, since the probability of malignancy is very low.

\section{Results}
\renewcommand{\arraystretch}{2}

\begin{table}[ht]
\centering
\caption{Validation scores of the compared models for significant prostate cancer detection. RS: Ranking Score, AUROC: Area Under the Receiver Operating Characteristic Curve, AP: Average Precision, M: Prostate gland mask, C: Cropped prediction}
\label{tab:my-table}
\begin{tabular}{ccccc}
\multicolumn{1}{l}{}                                  &                                              & \multicolumn{3}{c}{\textbf{Validation (100 Patients)}} \\ \cline{3-5} 
                      & \textbf{Experiments}       & \textbf{RS} & \textbf{AUROC} & \textbf{AP} \\ \hline
\multicolumn{1}{c|}{\multirow{3}{*}{\rotatebox[origin=c]{90}{\textbf{Single Model}}}}    & \textbf{nnU-Net semi}                         & 0.734               & 0.817               & 0.65                \\
\multicolumn{1}{c|}{} & \textbf{nnU-Net semi, M}    & 0.755       & 0.851          & 0.658       \\
\multicolumn{1}{c|}{} & \textbf{nnU-Net semi, M, C} & 0.77        & 0.864          & 0.677       \\ \hline
\hline
\multicolumn{1}{c|}{} & \textbf{nnU-Net semi}       & 0.712       & 0.821          & 0.602       \\
\multicolumn{1}{c|}{\multirow{4}{*}{\rotatebox[origin=c]{90}{\textbf{Ensemble Models}}}} & \textbf{nnU-Net semi, M}                      & 0.777               & 0.87                & 0.684               \\
\multicolumn{1}{c|}{} & \textbf{nnU-Net semi, M, C} & 0.807       & 0.885          & 0.729       \\
\multicolumn{1}{c|}{} & \textbf{nnU-Net semi, M, C, PSA th.}     & 0.81           & 0.888      & 0.732               \\
\multicolumn{1}{c|}{} & \textbf{nnU-Net semi, M, C, PSA th., ADC th.} & 0.81       & 0.888      & 0.732               \\ \hline

\end{tabular}
\end{table}


The prostate gland segmentation model was the first trained model. In this model, a 3D nnU-Net structure was implemented. 203 patients from the ProstateX dataset were used to train the 3D nnU-Net model in order to get probabilistic prostate gland segmentation. Only the T2 modality was considered in the prostate gland segmentation model since it offered the finest anatomical structure distinction among the T2, ADC, and DWI modalities. Five folds of the ProstateX dataset were created, and a model was trained for each fold. On the PICAI dataset of 1500 patients, prediction was done using a single model and an ensembled model to compare the performances. Since the ground truth prostate gland masks were not available for the 1500 patients of the PICAI dataset, it was not possible to determine the model performances directly; instead, the performances of the single model and the ensembled model were determined based on the differences in the predictions. Dice scores were calculated in order to compare the predictions of the single model and the predictions of the ensembled model. As a result, the dice scores obtained 95.6\% for the central zone and 92.4\% for the peripheral zone. Since the predictions of both models are fairly similar, it was decided to deploy only one model in order to cut down on workload and prediction time. Because the main objective of this model was to properly segment prostate zones, the loss function was set to Dice + CE loss during model training. 

\begin{figure}[htp]
    \centering
    \includegraphics[width=11 cm]{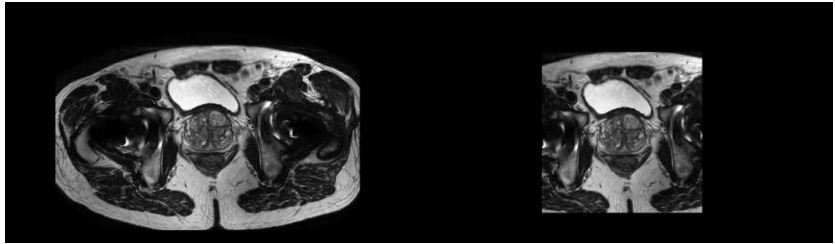}
    \caption{Sample image before and after crop}
    \label{fig:fig4}
\end{figure}

Fivefold dataset splits were predetermined for the experiments, and they persisted throughout all of the experiments. As a result, it was possible to correctly assess how training and postprocessing methods used in studies affected performances.  The first trained lesion segmentation model was the 3D nnU-Net model created as the base model (experiment 1). In this experiment, which was done as a single model, first fold data was included (1200 training, 300 validation) and T2, ADC and DWI modalities were used. In external validation (PICAI online validation with 100 patients), it received an area under the receiver operating characteristics curve (AUROC) of 0.817, an average precision (AP) of 0.65, and a ranking score (RS) of 0.734, which is the average of AUROC and AP. In the second experiment, the peripheral and central gland masks, which are the outputs of the prostate gland segmentation model, were fed into the model as two separate modalities in addition to T2, ADC, and DWI. In external validation, this model achieved AUROC of 0.851, AP of 0.658, and RS of 0.755. Based on these results, it was concluded that giving prostate masks as separate modalities to the model had a positive effect on performance. The second experiment's trained model was evaluated on cropped images in the third experiment, which did not include any model training. In this experiment, images having a wide field of view (FOV) were cropped, and it was attempted to eliminate as much of the regions from the images that were unlikely to contain lesions. The results of the validation on external dataset are: AUROC of 0.864, AP of 0.677, and RS of 0.77. Because the model predicted on smaller images rather than extremely broad images, the prediction result on cropped images may have been greater than that of earlier models. The Figure \ref{fig:fig4} shows a sample image before and after cropping. 

After the experiments carried out as a single model, ensemble experiments with 5 models were carried out. These experiments were started in order to prevent possible bias that may occur due to data splitting and to create stronger, more generalizable models by training 5 models for 5 folds by combining their power with ensemble. As a first of the ensemble experiments (experiment 4), 5 models were trained for 5 folds with T2, ADC and DWI modalities, so that a base model was created for the ensemble experiments, and it was possible to compare it with the single model trained with the same setup. The results of the validation on the external dataset are: AUROC of 0.861, AP of 0.602, and RS of 0.712. It was discovered that the single model performed better than the ensembled model when their respective performances were compared. The possible reason behind this was bias due to data distribution. In the next experiment (experiment 5), 5 models were trained for 5 folds with the dataset created by adding peripheral and central zones to T2, ADC and DWI modalities, and the performance of the ensemble model was tested on the external dataset. The resulting metrics are AUROC of 0.87, AP of 0.684 and RS of 0.777. These results highlight that prostate gland segmentations have a beneficial impact on performance in experiments applying ensemble models. Furthermore, the ensemble model outperformed the single model in the experiment with masks whereas the single model outperformed the ensemble model in the experiment without masks. These findings suggest that the performance that prostate gland masks add to the model allows the masked models to avoid biases brought on by distributions. The performance of ensemble models with prostate masks trained in Experiment 5 was assessed on cropped images in Experiment 6. The resulting metrics are RS of 0.807, AUROC of 0.885, and AP of 0.729. Predicting on cropped images has improved performance in ensemble models just as it did in single model.

\begin{figure}[htp]
    \centering
    \includegraphics[width=8 cm]{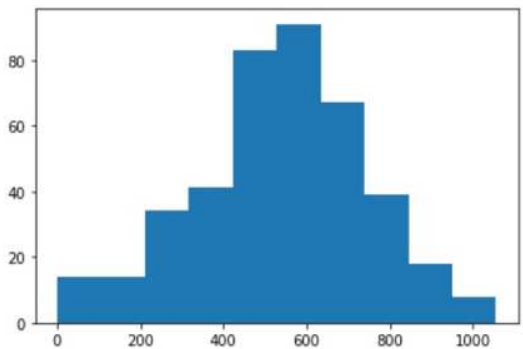}
    \caption{Histogram of ADC values}
    \label{fig:fig5}
\end{figure}

PSA thresholding was performed to analyze whether the model performance could be improved by combining the clinical data with the best-performing model (experiment 6) obtained as a result of the experiments with DL methods. Although a threshold of 0.1, 0.15, or 0.2 is considered appropriate in a previous study \cite{girometti2022comparison} for the differentiation of malignant and benign, in this experiment (experiment 7), 0.04 was chosen as the threshold value to prevent possible false negative predictions, therefore patients below this value were directly considered non-csPCA. The resulting metrics are RS of 0.81, AUROC of 0.888, and AP of 0.732. As the result shows, PSA thresholding did not significantly improve performance, but this was due to the threshold that was kept low to prevent possible false negatives. It should also be noted that validation was performed on a small dataset of 100 patients. In the last experiment, ADC intensity thresholding was applied. First, the ADC intensity distributions of the lesion areas of the patients with csPCA in the train data were found and a histogram was created by taking the average of the 10\% percentiles of those distributions. The histogram is as shown in the Figure \ref{fig:fig5}. The threshold value determined according to this histogram is 1200, this is because margin is added to prevent possible false negatives. Patients who exceeded this threshold were directly deemed to not have csPCA. The resulting metrics are RS of 0.81, AUROC of 0.888, and AP of 0.732. From what can be understood, there is no difference between the previous experiment. All results are summarized in Table \ref{tab:my-table}. The resulting metrics in the " Open Development Phase - Testing" of the best model obtained in the  "Open Development Phase - Validation and Tuning" are as follows: RS of 0.752, AUROC of 0.889, and AP of 0.614. 

\section{Discussion}

We proposed a nnU-Net based model for clinically significant prostate cancer detection. We included two key components to our model design: 1) introduction of probabilistic prostate gland segmentation as input for lesion detection, 2) Use of clinical markers such as PSAd and ADC lesion histograms to eliminate false positive predictions. Our proposed framework achieved top performance in open validation phase of PICAI challenge. One limitation of our approach is the use of clinical indices as a post-processing step. We believe utilizing clinically critical information as an additional input to the model can increase the clinically significant lesion detection similar to the daily clinical practice of the radiologist. In conclusion, we proposed a framework to detect clinically significant lesions on biparametric MRI, which can aid in avoiding contrast agent usage in current clinical setups.



\section*{Acknowledgements}
This paper has been produced benefiting from the 1001 Science and Technology Grant Program National Program of TUBITAK (Project No: 122E022). This paper has been produced benefiting from the 2232 International Fellowship for Outstanding Researchers Program of TUBITAK (Project No: 118C353). However, the entire responsibility of the publication/paper belongs to the owner of the paper. The financial support received from TUBITAK does not mean that the content of the publication is approved in a scientific sense by TUBITAK.
%
%
%
\bibliographystyle{splncs04}
\bibliography{main_picai_bib.bib}

\end{document}